\documentclass[aps,preprint,showpacs, showkeys]{revtex4}%
\usepackage{amsfonts}
\usepackage{amsmath}
\usepackage{amssymb}
\usepackage{graphicx}%
\providecommand{\U}[1]{\protect\rule{.1in}{.1in}}

\begin{document}
\title[ ]{Hermiticity Breaking and Restoration in the ($g\phi^{4}+h\phi^{6}$)$_{1+1}$
Field Theoretic Model}
\author{Abouzeid. M. Shalaby\footnote{E-mail:amshalab@ mans.eg.edu}}
\affiliation{Centre for Theoretical Physics, The British University in Egypt, El Sherouk
City, Misr Ismalia Desert Road, Postal No. 11837, P.O. Box 43, Egypt }
\affiliation{Physics Department, Faculty of Science, Mansoura University, Egypt.}

\begin{abstract}
We introduce hermiticity as a new symmetry and show that when starting with a
model which is Hermitian in the classical level, quantum corrections can break
hermiticity while the theory stay physically acceptable. To show this, we
calculated the effective potential of the ($g\phi^{4}+h\phi^{6}$)$_{1+1}$
model up to first order in $g$ and $h$ couplings which is sufficient as the
region of interest has finite correlation length for which mean field
calculation may suffice. We show that, in the literature, there is a skipped
phase of the theory due to the wrong believe that the theory in the broken
hermiticity phase is unphysical. However, in view of recent discoveries of the
reality of the spectrum of the non-Hermitian but $PT$ symmetric theories, in the
broken hermiticity phase the theory possesses  $PT$ symmetry and thus physically acceptable. In
fact, ignoring this phase will lead to violation of universality when
comparing this model predictions with other models in the same class of
universality.
\end{abstract}
\maketitle

Since the discovery of the reality of the eigen values of a class of
non-Hermitian Hamiltonian models ($PT$ symmetric)
\cite{bendg1,bend1,bend2,bend3,bend4,bend5,bend6,bend7,bend8,bend9,bend10,bend11,bend12,bend13,bend14}
and efforts are growing to solve the problems concerning their statistical
interpretation ( they have negative norm) and unitarity of the time evolution
operator. The solution of such problems can be achieved via two approaches.
The first approach, due to Bender $et.al$ , is to replace the bra in the Dirac
convention by a $CPT$ bra such that the new inner product takes the form
\cite{bend2005}
\[
\langle A|B\rangle_{CPT}=(CPT|A\rangle)^{T}|B\rangle,
\]
which replaces the conventional Dirac bracket $\ \langle A|B\rangle
=|A\rangle^{\dagger}|B\rangle.$ The operator $C$ is determined dynamically and
in most cases perturbatively. This approach succeeded in correcting the the
negative norm problem and the unitarity of the time evolution operator and
thus leads to the conservation of the probability density.

The other approach, due to A. Mostafazadeh \cite{zadah}, searches for a
similarity transformation which has the job of transforming the non-Hermitian
Hamiltonian operator $H$ into a Hermitian operator $h$, where $h=\eta
^{-1}H\eta.$ Again, the determination of the operator $\eta$ can be obtained
perturbatively in most cases.

After the solution of the most challenging problems in non-Hermitian theories
( ghosts and unitarity), \ people are trying to revisit old models which
previously was rejected. For instance, a class of simple but non-trivial
quantum mechanical models are given by%

\begin{equation}
H=p^{2}+x^{2}\left(  ix\right)  ^{\epsilon}\text{, \ \ \ }\epsilon>0\text{.}
\label{qwang}%
\end{equation}
All such models have real and positive spectra even in the case of
$\epsilon=2$. In fact, all the complex PT symmetric Hamiltonians are believed
to have real and positive spectra \cite{bend1}. For such theories to be
acceptable, an inner product is redefined in such a way to overcome the
existence of ghost states ( states with negative norm). For a complex
non-Hermitian Hamiltonian having an unbroken $PT$ symmetry, a linear operator
$C$ that commutes with both $H$ and $PT$ can be constructed. The inner product
with respect to $CPT$ conjugation satisfies the requirements for the theory
defined by $H$ to have a Hilbert space with a positive norm and to be a
consistent unitary theory of quantum mechanics.

Another model for which $PT$ symmetric non-Hermitian formulation saved its
validity is the Lee model \ which was introduced in the 1950s as an elementary
quantum field theory in which mass, wave function, and charge renormalization
could be carried out exactly. In early studies of this model it was found that
there is a critical value of $g^{2}$, the square of the renormalized coupling
constant, above which $g_{0}^{2}$, the square of the unrenormalized coupling
constant, is negative. Thus, for $g^{2}$ larger than this critical value, the
Hamiltonian of the Lee model becomes non-Hermitian. It was also discovered
that in this non-Hermitian regime a new state appears whose norm is negative.
This state is called a ghost state. It has always been assumed that in this
ghost regime the Lee model is an unacceptable quantum theory because unitarity
appears to be violated. However, in this regime while the Hamiltonian is not
Hermitian, it does possess PT symmetry. \ Again, the construction of an inner
product via the construction of a linear operator $C$ saves the theory from
physical unacceptability \cite{lee}.

\begin{figure}[ptb]
\begin{center}
\includegraphics{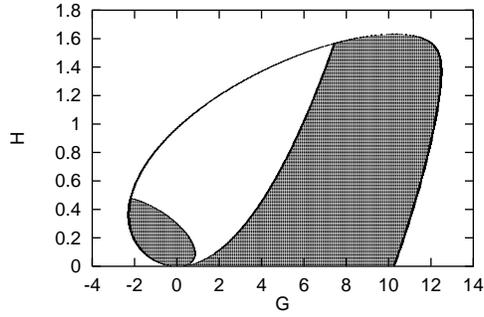}
\end{center}
\caption{The domain of symmetric $b=0$ solutions in $G$, $H$ parameter space
(shadowed area) and the broken $Z_2$ ( $b\neq 0$ and imaginary) and $H$ symmetry phases (the loop) taken from Ref. \cite{orpap} where $G=g/m^{2}$ and $H=h/m^{2}$. Outside the shadowed as well as the loop areas, $Z_2$ is broken ($b\neq 0$) while $H$ symmetry is restored. }%
\label{diagr}%
\end{figure}

We mentioned above two different algorithms by which non-Hermiticity appears. In the first one ( Eq.(\ref{qwang})), we started by a classical non-Hermitian models and mentioned that they are  physically acceptable. In the second one, the Lee model, the quantum corrections breaks the Hermiticity down while the system stays physically acceptable. In view of these two examples, one may ask if Hermiticity itself can be taken as a symmetry which can be broken or restored by the quantum corrections?.  
In this letter, we try to answer this question via the  investigation of the phase structure of the $\left(  g\phi^{4}+h\phi^{6}\right)  $ scalar field model which can show up both breaking and restoration of the Hermiticity due to quantum corrections. This model is believed to be Hermitian however, we will show that this is not the case for some couplings values. For these values, the $\left(  g\phi^{4}+h\phi^{6}\right)  $ model has  similar problems
concerning non-hermiticity and negative norm sates, wich can be cured through the prediction of the $C$ operator which is out of the scope of this letter .\ The authors of
Ref.\cite{orpap} named the ghost solutions of this model as spurus solutions
because in some regions of the phase plan ( Fig.\ref{diagr}) the Hamiltonian becomes non-Hermitian. Again, as in the early studies of the Lee model, it is claimed
that the theory in the regime of spurus solutions is physically unacceptable.
However, although the theory is non-Hermitian in the presence of spurus
solutions, it is $PT$ symmetric and thus represent a physically acceptable
model. In fact, ignoring some phases of this theory has it's draw back on the
conclusions drawn to chiral symmetry phase transition. This is because the
$\left(  g\phi^{4}+h\phi^{6}\right)  $ theory is believed to be in the same
class of universality with 3 flavors QCD near the Chiral phase transition
\cite{wilczk}. Thus it is certainly important to investigate carefully the
content of the phase structure of this model supported by the new formulation
of quantum theory which respects $PT$ symmetry in the same way we respected
hermiticity before.

 Now, consider the Hamiltonian density, normal-ordered\ with respect to
the mass $m$;
\begin{equation}
H=N_{m}\left(  \frac{1}{2}\left(  \left(  \triangledown\phi\right)  ^{2}%
+\pi^{2}+m^{2}\phi^{2}\right)  +g\phi^{4}+h\phi^{6}\right)  .\label{h6}%
\end{equation}
Here, we introduce a new symmetry into physics. The above model is invariant
under the operation \ $H\rightarrow H^{\dagger}$. \ Let us write Eq.(\ref{h6})
in a normal-ordered\ form with respect to the mass $M=t\cdot m$, using the
following relations:
\begin{align*}
N_{m}\Psi &  =N_{M}\Psi,\\
N_{m}\Psi^{2} &  =N_{M}\Psi^{2}+\Delta,\\
N_{m}\Psi^{3} &  =N_{M}\Psi^{3}+3\Delta N_{M}\Psi,\\
N_{m}\Psi^{4} &  =N_{M}\Psi^{4}+6\Delta N_{M}\Psi^{2}+3\Delta^{2},\\
N_{m}\psi^{5} &  =N_{M}\psi^{5}+10\Delta N_{M}\psi^{3}+15\Delta^{2}\psi,\\
N_{m}\psi^{6} &  =N_{M}\psi^{6}+15\Delta N_{M}\psi^{4}+45\Delta^{2}\psi
^{2}+15\Delta^{3}.
\end{align*}
Accordingly, after the application of the canonical transformation
\begin{equation}
\left(  \phi,\pi\right)  \rightarrow\left(  \psi+B,\Pi\right)  ,
\end{equation}
we can write the Hamiltonian as
\begin{equation}
H=\bar{H}_{0}+\bar{H}_{I}+\bar{H}_{1}+E,\label{hpt}%
\end{equation}
where%

\begin{align}
\bar{H}_{0} &  =N_{M}\left(  \frac{1}{2}\left(  \Pi^{2}+\left(  \triangledown
\psi\right)  ^{2}\right)  +\frac{1}{2}M^{2}\psi^{2}\right)  ,\\
\bar{H}_{I} &  =gN_{M}\left(  \psi^{4}+4B\psi^{3}\right)  \\
&  +hN_{M}\left(  \psi^{6}+6B\psi^{5}+\left(  15\Delta+15B^{2}\right)
\psi^{4}+\left(  60B\Delta+20B^{3}\right)  \psi^{3}\right)
\end{align}
Also%

\begin{align*}
\bar{H}_{1} &  =\frac{1}{2}\left(  m^{2}-M^{2}+12g\left(  B^{2}+\Delta\right)
+30h\left(  B^{4}+6\Delta B^{2}+3\Delta^{2}\right)  \right)  \psi^{2}\\
&  +\left(  m^{2}+4g\left(  B^{2}+3\Delta\right)  +6h\left(  B^{4}+10\Delta
B^{2}+15\Delta^{2}\right)  \right)  B\psi,
\end{align*}
and%

\begin{align}
E  &  =\frac{1}{2}\left(  m^{2}+12g\Delta\right)  B^{2}+\left(  g+15h\Delta
\right)  B^{4}+15h\left(  3\Delta B^{2}+\Delta^{2}\right)  \Delta\\
&  +hB^{6}+\frac{1}{8\pi}\left(  M^{2}-m^{2}\right)  +3g\Delta^{2}+\frac{1}%
{2}m^{2}\Delta.
\end{align}
Since $E$ serves as the generating \ functional for all the 1PI amplitudes, it
satisfies the renormalization conditions given by \cite{Peskin}%

\begin{equation}
\frac{\partial^{n}}{\partial b^{n}}E(b,t,G)=g_{n}\text{,}%
\end{equation}
where $g_{n}$ are the $\psi^{n}$ coupling. For instance,%
\begin{equation}
\frac{\partial E}{\partial B}=0\text{, \ \ \ }\frac{\partial^{2}E}{\partial
B^{2}}=M^{2}\text{,} \label{ren}%
\end{equation}
The first condition enforces $\bar{H}_{1}$ to be zero and thus
\begin{align*}
\frac{1}{2}\left(  m^{2}-M^{2}+12g\left(  B^{2}+\Delta\right)  +30h\left(
B^{4}+6\Delta B^{2}+3\Delta^{2}\right)  \right)   &  =0,\\
\left(  m^{2}+4g\left(  B^{2}+3\Delta\right)  +6h\left(  B^{4}++10\Delta
B^{2}+15\Delta^{2}\right)  \right)  B  &  =0.
\end{align*}
The use of the dimensionless parameters \ $t=\frac{M^{2}}{m^{2}}$, $G=\frac
{g}{4\pi m^{2}}$, $H=\frac{h}{\left(  4\pi m\right)  ^{2}}$ and $b^{2}=4\pi
B^{2}$, leads to the following results
\begin{align}
\left(  1-t+12G\left(  b^{2}-\ln t\right)  +30H\left(  b^{4}-6b^{2}\ln
t+3\left(  \ln t\right)  ^{2}\right)  \right)   &  =0,\nonumber\\
\left(  1+4G\left(  b^{2}-3\ln t\right)  +6H\left(  b^{4}-10b^{2}\ln
t^{2}+15\left(  \ln t\right)  ^{2}\right)  \right)  b  &  =0. \label{orphi6}%
\end{align}
Also the energy density takes the form
\begin{equation}
E=\frac{m^{2}}{8\pi}\left\{
\begin{array}
[c]{c}%
1+G\left(  b^{4}-12b^{2}\ln t+3G\left(  \ln t\right)  ^{2}\right)  -15H\ln
t\left(  b^{4}-3b^{2}\ln t+\left(  \ln t\right)  ^{2}\right) \\
+Hb^{6}+\frac{1}{8\pi}\left(  t-1-\ln t\right)
\end{array}
\right\}
\end{equation}
For some specific values of $G$ and $H$, one solves Eq.(\ref{orphi6}) to get
the value of $b$ and $t$. Thus, as $t$ chosen to be greater than zero, the
solutions determine the parameters at the minima of the energy density.

Fortunately, the equation set in Eq.(\ref{orphi6}) coincides with the
prediction of GEP calculations in Ref.\cite{gep6} and the Oscillator Representation method in Ref.\cite{orpap}. The region of applicability of the
theory in terms of the parameters $G$ and $\ H$ is analyzed in
Ref.\cite{gep6} with the $PT$ symmetric phase is totally ignored while in
Ref.\cite{orpap} the authors claimed that it is unphysical. In fact, although
the Hamiltonian is non-Hermitian in the phase of imaginary condensate it does
possess a $PT$ symmetry and thus having a real spectrum.

For regions in the coupling space far a way from the second order phase
transition, it is well known that the theory has a finite correlation length ( non-zero mass parameter).
Accordingly, even mean field calculations suffices to describe the theory
\cite{wilczk} for the region around the $PT$ symmetric phase. Thus, the equations obtained from normal-ordered effective potential ( it is up to first order quantum corrections in $G$ and $H$) \ can be considered reliable for the region of interest.

By solving Eq(\ref{orphi6}), one can obtain the phase diagram of the theory
\cite{orpap}. In Fig.\ref{diagr}, we realize that the theory possesses $Z_{2}$
and $H$ ( hermiticity) symmetries in the shadowed area. In the loop sandwiched
in the shadowed area, both symmetries are broken ( imaginary condensate).
However, if we look at the Hamiltonian form in Eq.(\ref{hpt}), imaginary
condensates turns the theory non-Hermitian and $PT$ symmetric as well. Accordingly,
the theory in this region is physically acceptable as it is believed to have
real eigen values. In fact, in our calculations, the first state in the
spectrum ( the effective potential) is real and thus agree with the Bender's
conjecture that $PT$ symmetric theories have real eigen values. Outside the
shadowed area and the loop, the $H$ symmetry is restored and $Z_{2}$ stay broken.

For the calculations of different amplitudes which represent physical
quantities one has to resort to Benders convention for the inner products
which in turn demands the calculation of $C$ operator for the theory under
investigation and try to find the new Feynman rules. In fact, this will teak a substantial amount of time and it naturally becomes a topic of our future work. 

In conclusion, we introduced a novel symmetry in physics, the $H$ symmetry,
and showed that it can be broken and restored due to the quantum corrections.
Also, among the novelty of this letter, we showed that the $\left(
g\phi^{4}+h\phi^{6}\right)$ theory has a
previously missed phase which we explain that it is physically acceptable as
the theory is $PT$ symmetric though it has imaginary condensate and thus
non-Hermitian. The importance of exploring this phase is that, the $\left(
g\phi^{4}+h\phi^{6}\right)  $ theory is believed to be in the same class of
universality with 3 flavors QCD near the Chiral phase transition and without
that phase wrong conclusions may be drawn concerning \ universality classes.

\newpage

\newpage


\begin{thebibliography}{99}                                                                                               %


\bibitem {bendg1}Carl M. Bender, Peter N. Meisinger, and Haitang Yang, Phys.
Rev. D63, 045001 (2001).

\bibitem {bend1}C. M. Bender, S. Boettcher, and P. N. Meisinger, J. Math.
Phys. 40, 2201 (1999).

\bibitem {bend2}C. M. Bender and S. Boettcher, Phys. Rev. Lett. 80, 5243 (1998).

\bibitem {bend3}C. M. Bender, F. Cooper, P. N. Meisinger, and V. M. Savage,
Phys. Lett. A 259, 224 (1999).

\bibitem {bend4}C. M. Bender and G. V. Dunne, J. Math. Phys. 40, 4616 (1999).

\bibitem {bend5}E. Delabaere and F. Pham, Phys. Lett. A 250, 25 (1998) and 29 (1998).

\bibitem {bend6}E. Delabaere and D. T. Trinh, J. Phys. A: Math. Gen. 33, 8771 (2000).

\bibitem {bend7}G. A. Mezincescu, J. Phys. A: Math. Gen. 33, 4911 (2000).

\bibitem {bend8}C. M. Bender, S. Boettcher, and V. M. Savage, J. Math. Phys.
41, 6381 (2000).

\bibitem {bend9}C. M. Bender and Q. Wang, J.Phys. A34, 9835 (2001).

\bibitem {bend10}K. C. Shin, University of Illinois preprint.

\bibitem {bend11}C. M. Bender and E. J. Weniger, J. Math. Phys. 42, 2167 (2001).

\bibitem {bend12}C. M. Bender, G. V. Dunne, P. N. Meisinger, and M.Simsek,
Phys. Lett. A 281, 311 (2001).

\bibitem {bend13}P. Dorey, C. Dunning, and R. Tateo, hep-th/0103051.

\bibitem {bend14}C. M. Bender, M. Berry, P. N. Meisinger, V. M. Savage, and M. S.ims.

ek, J. Phys. A:Math. Gen. 34, L31-L36 (2001).

\bibitem {bend2005}Carl M. Bender, Jun-Hua Chen, Kimball A. Milton,
J.Phys.A39, 1657 (2006).

\bibitem {zadah}A. Mostafazadeh, J. Phys. A: Math. Gen. 38, 6557 (2005) and
Erratum 38, 8185 (2005).

\bibitem {orpap}A.V.Vinikov, C.R.Ji, J.I.Kim and D.P. Min, arXiv:hep-ph/0204114v2.

\bibitem {wilczk}Frank Wilczek, arXiv:hep-ph/0003183 v1 (2000).

\bibitem {Peskin}Michael E.Peskin and Daniel V.Schroeder, AN INTRODUCTION TO
THE QUANTUM FIELD THEORY (Addison-Wesley Advanced Book Program, 1995).

\bibitem {gep6}P.~M.~Stevenson and I.~Roditi,
Phys.\ Rev.\ D \textbf{33}, 2305 (1986).


\bibitem {lee}Carl M. Bender, Sebastian F. Brandt, Jun-Hua Chen, Qing-hai
Wang, \ Phys.Rev.D71, 025014 (2005).
\end{thebibliography}
\end{document}